\documentclass[aps,prc,preprint,groupedaddress,showpacs]{revtex4}
\usepackage{epsfig}
\usepackage{amsmath}

\begin{document}

\preprint{OSU/204-TKE}

\title{The Excitation Energy Dependence of the Total Kinetic Energy Release in $^{235}$U(n,f)}

\author{R. Yanez}
\author{L. Yao}
\author{J. King}
\author{W. Loveland}
\affiliation{Department of Chemistry, Oregon State University,
 Corvallis, OR, 97331.}
 \author{F. Tovesson}
 \author{N. Fotiades}

\affiliation{Los Alamos National Laboratory, Los Alamos, NM 87545 }

\date{\today}

\begin{abstract}
The total kinetic energy release in the neutron induced fission of $^{235}$U was
measured (using white spectrum neutrons from LANSCE) for neutron energies from
E$_{n}$ = 3.2 to 50 MeV.   In this energy range the average post-neutron total kinetic energy release drops from 167.4 $\pm$ 0.7 to 162.1 $\pm$ 0.8 MeV, exhibiting a local dip near the second chance fission threshold.  The values  and the slope of the TKE vs. E$_{n}$ agree with previous measurements  but do disagree (in magnitude) with systematics.  The variances of the TKE distributions are larger than expected and apart from structure near the second chance fission threshold, are invariant for the neutron energy range from 11 to 50 MeV.  We also report the dependence of the total excitation energy in fission, TXE, on neutron energy.

\end{abstract}


\maketitle

\section{Introduction}
Most of the energy released in the nuclear fission process appears in the kinetic energy of the fission fragments.
  A first order estimate of the magnitude of the total kinetic energy release is that of the Coulomb energy of the fragments at scission, i.e., 
\begin{equation}
V_{Coul}=\frac{Z_{1}Z_{2}e^{2}}{r_{1}+r_{2}}
\end{equation}
where Z$_{n}$, r$_{n}$ are the atomic numbers and radii of fragments 1 and 2.  Recognizing that the fragments are deformed at scission, one can re-write equation 1 as 
\begin{equation}
TKE=\frac{Z_{1}Z_{2}e^{2}}{1.9(A_{1}^{1/3}+A_{2}^{1/3})}
\end{equation}
where the coefficient 1.9 (instead of the usual 1.2 - 1.3) represents the fragment deformation.  For symmetric fission, Z$_{1}$=Z$_{2}$=Z/2 and A$_{1}$ =A$_{2}$=A/2, then we have
\begin{equation}
TKE = (0.119)\frac{Z^{2}}{A^{1/3}}MeV
\end{equation}
Trajectory calculations \cite{raja} for alpha particle emission in fission have shown that the fission
 fragments are in motion at scission with a pre-scission kinetic energy of 7.3 MeV and an additive term representing this motion is needed. 
 Thus we have the ``Viola systematics" \cite{vic} that say 
\begin{equation}
TKE = (0.1189\pm 0.0011)\frac{Z^{2}}{A^{1/3}}+7.3(\pm 1.5)MeV
\end{equation}

The deformed scission point fragments will contract to their equilibrium deformations and the energy
 stored in deformation will be converted into internal excitation energy.  Thus we can define a related quantity,  the total excitation energy , TXE, in fission as
\begin{equation}
TXE=Q-TKE
\end{equation}where Q is the mass-energy release.  One quickly realizes that these quantities depend
 on the mass split in fission which in turn, at low excitation energies, may reflect the fragment nuclear structure.  The TXE is the starting point for calculations of the prompt neutron and gamma emission in fission, the yields of beta emitting fission fragments, reactor anti-neutrino spectra, etc.  As such, it is a fundamental property of all fissioning systems and sadly not very well known.

As a practical matter, one needs to know the dependence of the TKE and TXE on neutron 
energy for the neutron induced fission of technologically important actinide fissioning systems
 like $^{233}$U(n,f),$^{235}$U(n,f), and $^{239}$Pu(n,f).  The first question we might pose is
 whether the TKE should depend on the excitation energy of the fissioning system. 
 Does the energy brought in by an incident neutron in neutron induced fission appear
 in the fragment excitation energy or does it appear in the total kinetic energy? 
 In a variety of experiments, one finds that increasing the excitation energy of the 
fissioning system does not lead to significant increases in the TKE of the fission 
fragments or changes in the fragment separation at scission. \cite{VH}.  However,
 there may be more subtle effects that render this statement false in some circumstances. 
 For example, we expect, on the basis of the Coulomb energy systematics given above, 
 that the TKE will be proportional to changes in the fission mass splits which in turn can depend on the excitation energy.
    
For the technologically important reaction $^{235}$U(n,f), Madland \cite{dave} summarizes the known data \cite{straede, meadows, muller}with the following equations
\begin{equation}
\left\langle T_{f}^{tot}\right\rangle =\left( 170.93\pm 0.07\right) -\left(
0.1544\pm 0.02\right) E_{n}(MeV)
\end{equation}
\begin{equation}
\left\langle T_{p}^{tot}\right\rangle =\left( 169.13\pm 0.07\right) -\left(
0.2660\pm 0.02\right) E_{n}(MeV) 
\end{equation}
where E$_{n}$ is the energy of the incident neutron and T$_{f}^{tot}$and T$_{p}^{tot}$ are the
 average total fission fragment kinetic energy (before neutron emission) and the average fission
 product kinetic energy after neutron emission, respectively.  These quantities are related by the relation
\begin{equation}
\left\langle T_{p}^{tot}(E_{n}\right\rangle =\left\langle
T_{f}^{tot}(E_{n}\right\rangle \left[ 1-\frac{\overline{\nu _{p}}(E_{n})}{2A}%
\left( \frac{\left\langle A_{H}\right\rangle }{\left\langle
A_{L}\right\rangle }+\frac{\left\langle A_{L}\right\rangle }{\left\langle
A_{H}\right\rangle }\right) \right] 
\end{equation}
These data show a modest decrease in TKE with increasing excitation energy for the neutron
 energy interval E$_{n}$ =1-9 MeV.  There is no clearly identified changes in the TKE values
 near the second chance fission threshold, a feature that is important in semi-empirical models
 of fission such as represented by the GEF code.\cite{khs}  

In this paper, we report the results of measuring the total kinetic energy release in the neutron 
induced fission of $^{235}$U for neutron energies E$_{n}$ = 3.2 -50  MeV.  The method used for the 
measurement is the 2E method, i.e., measurement of the kinetic energies of the two coincident fission 
products using semiconductor detectors.  The time of flight of the neutrons inducing fission was measured, 
allowing deduction of their energy.  The details of the experiment are discussed in Section II while the 
experimental results and a comparison of the results 
with various models and theories is made in Section III with conclusions being summarized in Section IV.

\section{Experimental}

This experiment was carried out at the Weapons Neutron Research Facility (WNR) at the Los Alamos Neutron Science Center (LANSCE) at the Los Alamos  National Laboratory \cite{Lis, Liso}. ``White spectrum" neutron beams were generated from an unmoderated tungsten spallation source using the 800 MeV proton beam from the LANSCE linac.  The  experiment was located on the 15R  beam line (15$^{\circ}$-right with respect to the proton beam).  The calculated (MCNPX) ``white spectrum " at the target position is shown in figure 1.  \cite{snow} The proton beam is pulsed allowing one to measure the time of flight (energy) of the neutrons arriving at the experimental area.

A schematic diagram of the experimental apparatus is shown in figure 2. The neutron beam was collimated to a 1 cm diameter at the entrance to the experimental area. At the entrance to the scattering chamber, the beam diameter was measured to be 1.3 cm. A fission ionization chamber \cite{steve} was used to continuously monitor the absolute neutron beam intensities. The $^{235}$U target and the Si PIN diode fission detectors were housed in an  evacuated, thin-walled aluminum scattering chamber. The scattering chamber was located $\sim$ 3.1 m from the collimator, and $\sim$ 11 m from the neutron beam dump.  The center of the scattering chamber was located 16.46 m from the production target.

The $^{235}$U target consisted of a deposit of $^{235}$UF$_{4}$ on a thin C backing.  The thickness of the $^{235}$U was 175.5 $\mu$g $^{235}$U/cm$^{2}$ while the backing thickness was 100 $\mu$g/cm$^{2}$.  The isotopic purity of the $^{235}$U was 99.91 $\%$.  The target was tilted at 50 $^{\circ}$ with respect to the incident beam.

Fission fragments were detected by two arrays of Si PIN photodiodes (Hamamatsu S3590-09) arranged on opposite sides of the beam.  The area of the individual PIN diodes was 1 cm$^{2}$. The distance of the detectors from the target varied with angle from 2.60 cm to 4.12 cm.   The coincident detector pairs were at approximately 45, 60, 90, 115, and 135 $^{\circ}$.  The alpha particle energy resolution of the diodes was 18 keV for the 5475 keV line of $^{241}$Am. 

The time of flight of each interacting neutron was measured using a timing pulse from a Si PIN diode and the accelerator RF signal.  Absolute calibrations of this time scale were obtained from the photofission peak in the fission spectra and the known flight path geometry.  

The energy calibration of the fission detectors was done with a $^{252}$Cf source.  We have used the traditional Schmitt method  \cite{hal}.  Some have criticized this method especially for PIN diodes.  However with our limited selection of detectors, we were unable to apply the methods of \cite{moz} to achieve a robust substitute for the Schmitt method.

The measured fragment energies have be to be corrected for energy loss in the $^{235}$UF$_4$ deposit and the C backing foil. This correction was done by scaling the energy loss correction given by the Northcliffe-Schilling energy loss tables \cite{NS} to a measured mean energy loss of collimated beams of light and heavy $^{252}$Cf fission fragments in 100 $\mu$ g/cm$^{2}$ C foils. The scaling factor that was used was a linear function of mass using the average loss of the heavy and light fission fragments as anchor points. The correction factors at the anchor points were 1.24 and 1.45 for the heavy and light fragments, respectively. Similar factors were obtained if the SRIM code \cite{srim} was used to calculate dE/dx. These large deviation factors from measured to calculated fission fragment stopping powers have been observed in the past \cite{Knyazheva}, and represent the largest systematical uncertainty in the determination of the kinetic energies.

\section{Results and Discussion}

The measured average post-neutron emission fission product total kinetic energy release for the $^{235}$U(n,f) reaction(Table 1) is shown in Figure 3 along with other data and predictions
\cite{gunn, kapoor,  stevenson}.  The evaluated post-neutron emission data from Madlund \cite{dave} are shown as a dashed line while the individual pre-neutron emission measurements of \cite{muller} are shown as points.   The point at E$_{n}$ =14 MeV is the average of \cite{gunn} and \cite{stevenson}. The slope of the measured TKE release (this work) is in rough agreement with the previous measurements \cite{dave} at lower energies. Also shown are the predictions of the GEF model \cite{khs}.  GEF is a semi-empirical model of fission that provides a good description of fission observables using a modest number of adjustable parameters. The dashed line in Figure 1 is a semi-empirical equation (TKE = 171.5 -0.1E* for E* $>$ 9 MeV) suggested by Tudora et al. \cite{tudy} Qualitatively the decrease in TKE with increasing neutron energy reflects the increase in symmetric fission (with its lower associated TKE release) with increasing excitation energy.  This general dependence is reflected in the GEF code predictions with the slope of our data set being similar to  the predictions of the GEF model but with  the absolute values of the TKE release being  substantially less. 

In Figure 4, we show some typical TKE distributions along with Gaussian representations of the data.  In general, the TKE distributions appear to be Gaussian in shape.  This is in contrast to previous studies \cite{PR,D} which showed a sizable skewness in the distributions.

In Figure 5, we show the dependence of the measured values of the variance of the TKE distributions as a function of neutron energy along with the predictions of the GEF model of the same quantity. The measured variances are larger than expected.  
At low energies (near the second chance fission threshold) the observed variances show a dependence on neutron energy similar to that predicted by the GEF model, presumably reflecting the changes in variance with decreasing mass asymmetry.  At higher energies (11-50 MeV) the variances are roughly constant with changes in neutron energy.  Models \cite{poop} would suggest that most of the variance of the TKE distribution is due to fluctuations in the nascent fragment separation at scission.  The constancy of the variances  is puzzling.

Using the Q values predicted by the GEF code, one can make a related plot (Fig. 6) of the TXE values in the $^{235}$U(n,f) reaction.  The ``bump" in the TXE at lower neutron energies is pronounced and the dependence of the TXE upon neutron energy agrees with the GEF predictions although the absolute values are larger.

\section{Conclusions}

We conclude that : (a) For the first time, we have measured the TKE release and its variance for the technologically important $^{235}$U(n,f) reaction over a large range of neutron energies (3.2 - 50 MeV).  (b) The dependence of the TKE upon E$_{n}$ seems to agree with semi-empirical models although the absolute value does not.  (c) Understanding the variance and its energy dependence for the TKE distribution remains a challenge.

\begin{acknowledgments}

This work was supported in part by the
Director, Office of Energy Research, Division of Nuclear 
Physics of the Office of High Energy and Nuclear Physics 
of the U.S. Department of Energy
under Grant DE-FG06-97ER41026.  One of us (WL) wishes to  thank the [Department of Energy's]
 Institute for Nuclear Theory at the University of Washington for its hospitality
 and the Department of Energy for partial support during the completion of this work.
 This work has benefited from the use of the Los Alamos Neutron Science Center at the Los Alamos National Laboratory.  This facility is funded by the U. S. Department of Energy under DOE Contract No. DE-AC52-06NA25396.
 
 \end{acknowledgments}

\begin{table}[h]
\caption{Measured TKE release for $^{235}$U(n,f) }
\centering
\begin{tabular}{ccc}
\hline
E$_{n}$ (MeV)&$ \overline{TKE}$(MeV) & Uncertainty ($\overline{TKE}$)(MeV)\\
\hline
3.7&167.4&0.7\\
4.7&165.7&0.8\\
5.8&167.7&0.8\\
7.2&166.5&0.8\\
9.0&166.2&0.8\\
11.8&165.1&0.7\\
16.8&163.4&0.7\\
24.2&162.9&0.7\\
34.2&161.5&0.8\\
45.0&162.1&0.8\\
\hline
\end{tabular}
\end{table}

\newpage 
\begin{figure}[tbp]
\begin{minipage}{30pc}
\begin{center}
\includegraphics[width=100mm] {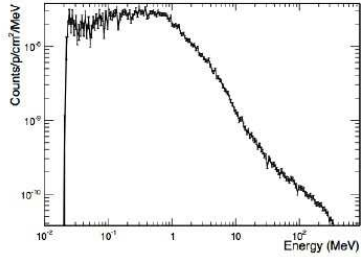}
\end{center}
\caption{The calculated neutron spectrum in the 15R beam area \cite{snow}}
\label{fig1}
\end{minipage}
\end{figure}

\newpage 
\begin{figure}[tbp]
\begin{minipage}{30pc}
\begin{center}
\includegraphics [width=100mm]{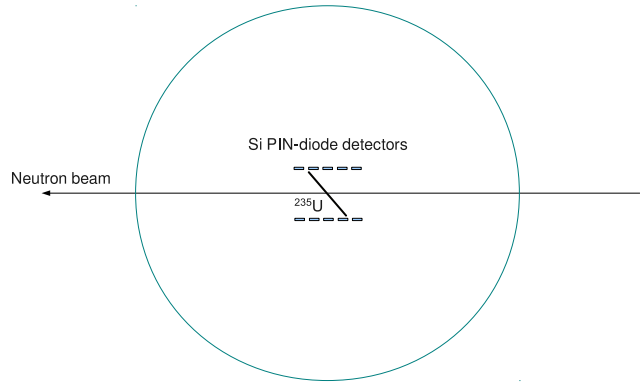}
\end{center}
\caption{(Color-online) Schematic diagram of the experimental apparatus. }
\label{fig2}
\end{minipage}
\end{figure}

\newpage 
\begin{figure}[tbp]
\begin{minipage}{30pc}
\begin{center}
\includegraphics [width=100mm]{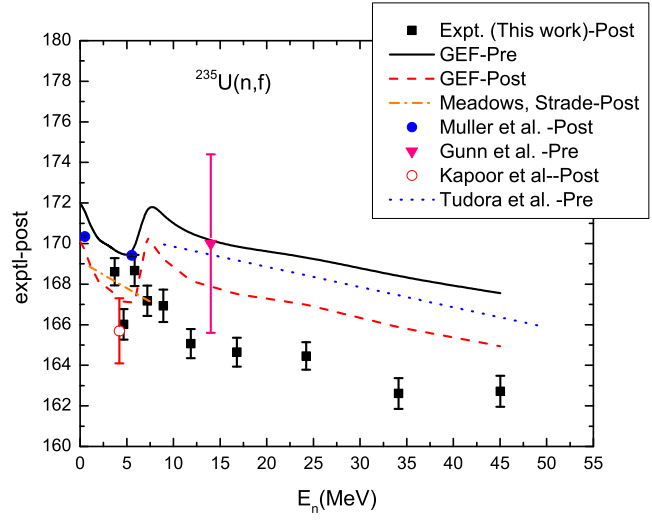}
\end{center}
\caption{(Color-online) TKE release data for $^{235}$U(n,f) }
\label{fig3}
\end{minipage}
\end{figure}

\newpage 
\begin{figure}[tbp]
\begin{minipage}{30pc}
\begin{center}
\includegraphics [width=100mm]{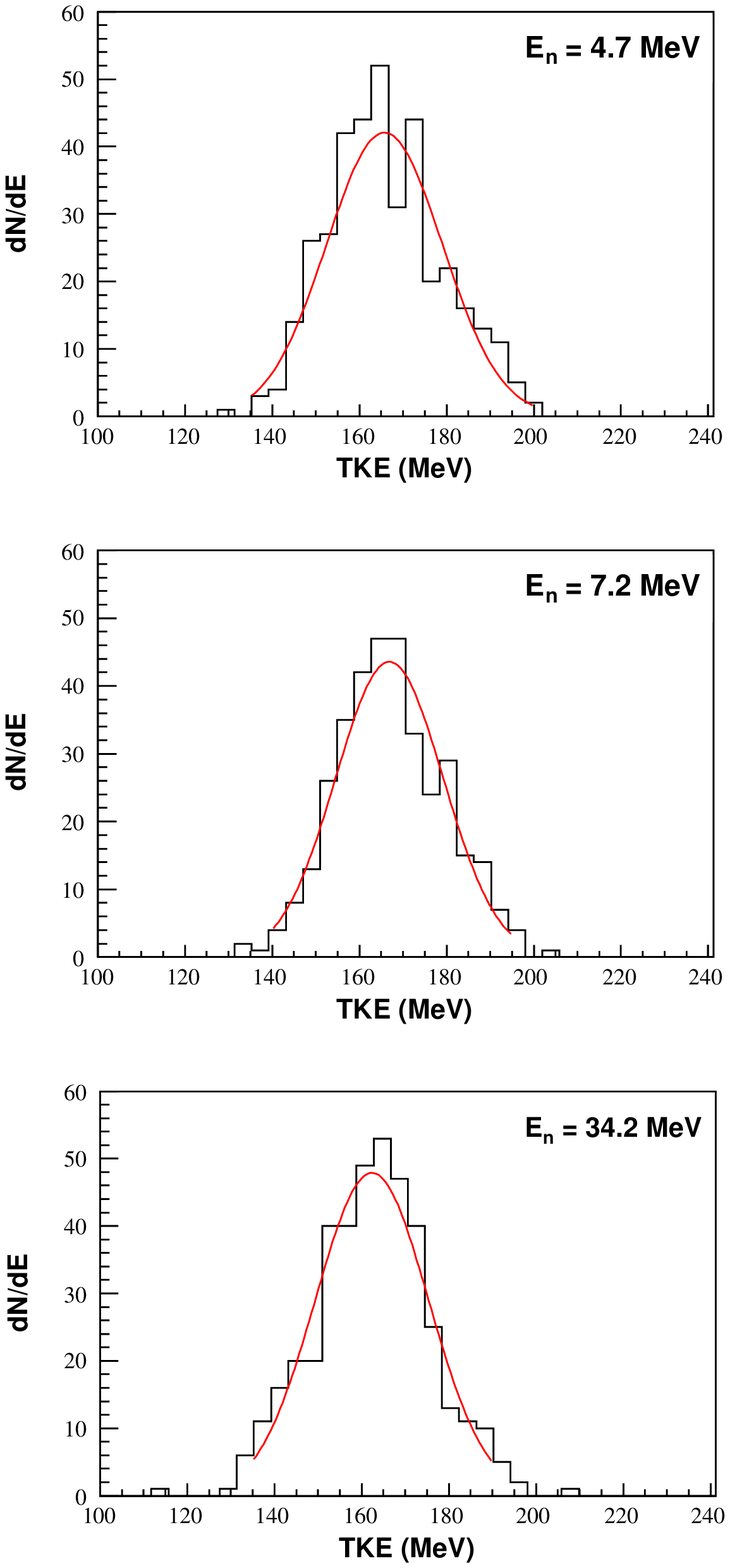}
\end{center}
\caption{(Color-online) Typical TKE distributions for $^{235}$U(n.f) }
\label{fig4}
\end{minipage}
\end{figure}

\newpage 
\begin{figure}[tbp]
\begin{minipage}{30pc}
\begin{center}
\includegraphics [width=100mm]{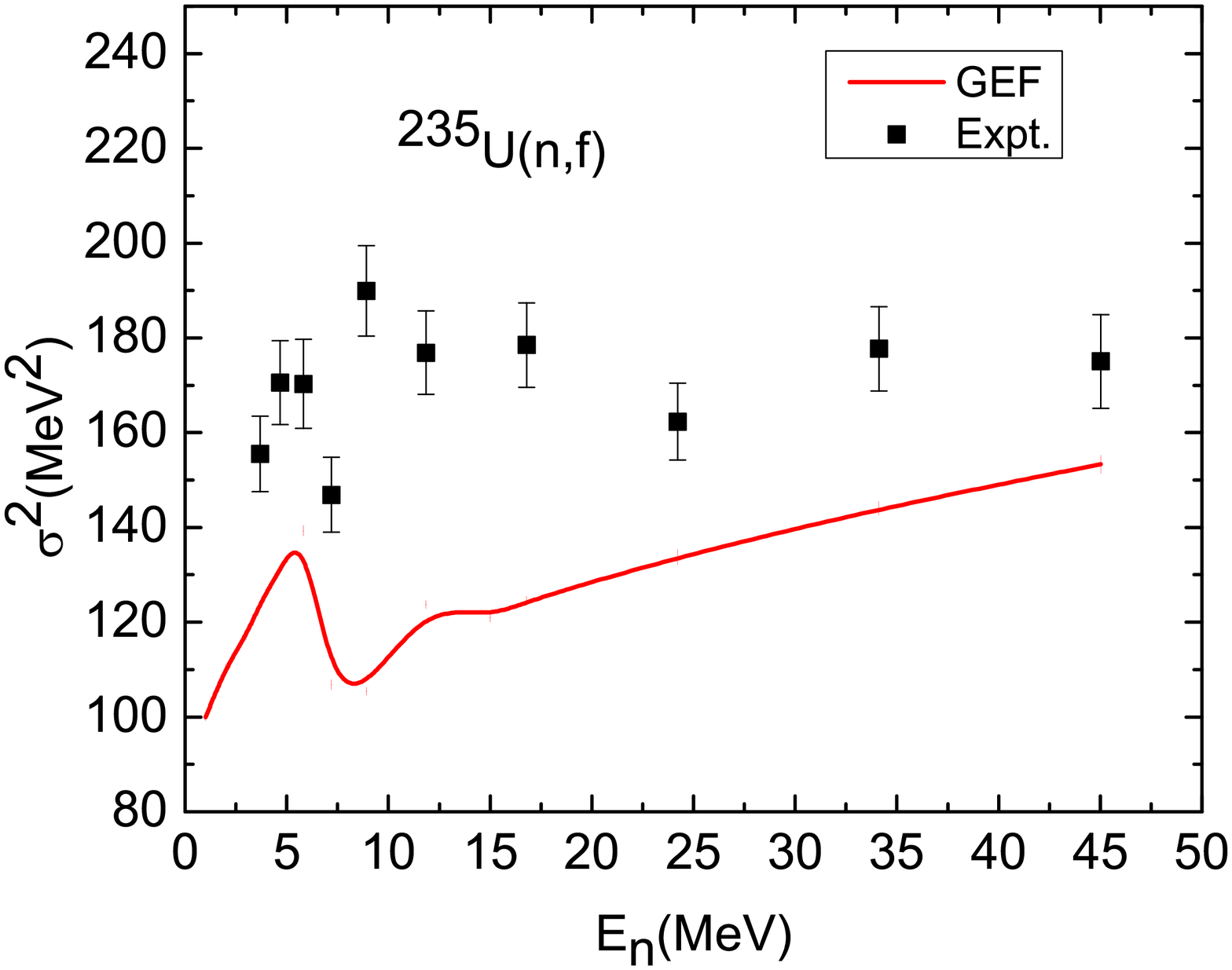}
\end{center}
\caption{(Color-online) Variance of the TKE distribution data for $^{235}$U(n,f) }
\label{fig5}
\end{minipage}
\end{figure}

\newpage 
\begin{figure}[tbp]
\begin{minipage}{30pc}
\begin{center}
\includegraphics [width=100mm]{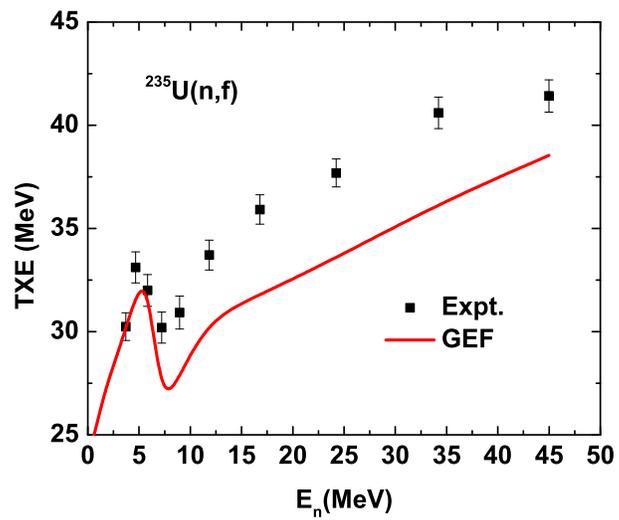}
\end{center}
\caption{(Color-online) TXE  data for $^{235}$U(n,f) }
\label{fig6}
\end{minipage}
\end{figure}

\end{document}